\documentclass[twocolumn,showpacs,amsmath,amssymb,prl]{revtex4}

\usepackage{graphicx}
\usepackage{nicefrac}
\usepackage{amsmath}
\usepackage{comment}
\usepackage{circuitikz}
\usepackage[all]{xy}
\usepackage{tikz}
\usepackage{amsmath}
\usepackage{textcomp}

\newcommand{\bra}[1]{\langle #1|}
\newcommand{\ket}[1]{|#1\rangle}
\newcommand{\braket}[2]{\langle #1|#2\rangle}

\newcommand{\ba}{\begin{array}}
\newcommand{\ea}{\end{array}}

\newcommand{\be}{\begin{align}}
\newcommand{\ee}{\end{align}}

\newcommand{\networkone}[1][]{
\begin{tikzpicture}[#1]
\tikzset{vertex/.style = {}}
\tikzset{edge/.style = {->,> = latex'}}
\node[vertex] (a) at  (0,0) {};
\node[vertex] (b) at  (1,0.833) {};
\node[vertex] (c) at  (2,0){} ;
\node[vertex] (d) at  (1,1.72) {};
\draw[edge] (b) to (a);
\draw[edge] (b) to (c);
\draw[edge] (b) to (d);
\draw[edge] (a) to (d);
\draw[edge] (d) to (c);
\draw[edge] (c) to (a);
\foreach \x in {(a), (b), (c), (d)}{ \fill \x circle[radius=2pt];}
\end{tikzpicture}
} 

\newcommand{\networktwo}[1][]{
\begin{tikzpicture}[#1]
\tikzset{vertex/.style = {}}
\tikzset{edge/.style = {-}}
\node[vertex] (a) at  (0,0) {};
\node[vertex] (b) at  (1,0.833) {};
\node[vertex] (c) at  (2,0) {};
\node[vertex] (d) at  (1,1.72) {};
\draw[edge] (b) to (a);
\draw[edge] (b) to (c);
\draw[edge] (b) to (d);
\draw[dashed] (a) to (d);
\draw[dashed] (d) to (c);
\draw[dashed] (c) to (a);
\foreach \x in {(a), (b), (c), (d)}{ \fill \x circle[radius=2pt];}
\end{tikzpicture}
}

\newcommand{\networkthree}[1][]{
\begin{tikzpicture}[#1]
\tikzset{vertex/.style = {}}
\tikzset{edge/.style = {-}}
\node[vertex] (a) at  (0,0) {};
\node[vertex] (b) at  (1,0.833) {};
\node[vertex] (c) at  (2,0) {};
\node[vertex] (d) at  (1,1.72) {};
\draw[edge] (b) to (a);
\draw[edge] (b) to (c);
\draw[dashed] (b) to (d);
\draw[edge] (a) to (d);
\draw[dashed] (d) to (c);
\draw[dashed] (c) to (a);
\foreach \x in {(a), (b), (c), (d)}{ \fill \x circle[radius=2pt];}
\end{tikzpicture}
}

\newcommand{\networkfour}[1][]{
\begin{tikzpicture}[#1]
\tikzset{vertex/.style = {}}
\tikzset{edge/.style = {->,> = latex'}}
\node[vertex] (a) at  (0,0) {};
\node[vertex] (b) at  (1,0.833) {};
\node[vertex] (c) at  (2,0) {};
\node[vertex] (d) at  (1,1.72) {};
\draw[edge] (b) to (a);
\draw[edge] (d) to (b);
\draw[edge] (a) to (d);
\foreach \x in {(a), (b), (d)}{ \fill \x circle[radius=2pt];}
\end{tikzpicture}
}

\newcommand{\networkfive}[1][]{
\begin{tikzpicture}[#1]
\tikzset{vertex/.style = {}}
\tikzset{edge/.style = {->,> = latex'}}
\node[vertex] (a) at  (0,0) {};
\node[vertex] (b) at  (1,0.833) {};
\node[vertex] (c) at  (2,0) {};
\node[vertex] (d) at  (1,1.72) {};
\draw[edge] (d) to (c);
\draw[edge] (c) to (b);
\draw[edge] (b) to (d);
\foreach \x in {(b), (c), (d)}{ \fill \x circle[radius=2pt];}
\end{tikzpicture}
}

\newcommand{\networksix}[1][]{
\begin{tikzpicture}[#1]
\tikzset{vertex/.style = {}}
\tikzset{edge/.style = {->,> = latex'}}
\node[vertex] (a) at  (0,0) {};
\node[vertex] (b) at  (1,0.833) {};
\node[vertex] (c) at  (2,0) {};
\node[vertex] (d) at  (1,1.72) {};
\draw[edge] (a) to (b);
\draw[edge] (b) to (c);
\draw[edge] (c) to (a);
\foreach \x in {(a), (b), (c)}{ \fill \x circle[radius=2pt];}
\end{tikzpicture} 
}

\newcommand{\networkseven}[1][]{
\begin{tikzpicture}[#1]
\tikzset{vertex/.style = {}}
\tikzset{edge/.style = {->,> = latex'}}
\node[vertex] (a) at  (0,0) {};
\node[vertex] (b) at  (1,0.833) {};
\node[vertex] (c) at  (2,0) {};
\node[vertex] (d) at  (1,1.72) {};
\draw[edge] (b) to (d);
\draw[edge] (a) to (d);
\draw[edge] (c) to (d);
\foreach \x in {(a), (b), (c), (d)}{ \fill \x circle[radius=2pt];}
\end{tikzpicture}
}

\newcommand{\networkeight}[1][]{
\begin{tikzpicture}[#1]
\tikzset{vertex/.style = {}}
\tikzset{edge/.style = {->,> = latex'}}
\node[vertex] (a) at  (0,0) {};
\node[vertex] (b) at  (1,0.833) {};
\node[vertex] (c) at  (2,0) {};
\node[vertex] (d) at  (1,1.72) {};
\draw[edge] (b) to (c);
\draw[edge] (a) to (c);
\draw[edge] (d) to (c);
\foreach \x in {(a), (b), (c), (d)}{ \fill \x circle[radius=2pt];}
\end{tikzpicture}
}

\newcommand{\networknine}[1][]{
\begin{tikzpicture}[#1]
\tikzset{vertex/.style = {}}
\tikzset{edge/.style = {->,> = latex'}}
\node[vertex] (a) at  (0,0) {};
\node[vertex] (b) at  (1,0.833) {};
\node[vertex] (c) at  (2,0) {};
\node[vertex] (d) at  (1,1.72) {};
\draw[edge] (b) to (a);
\draw[edge] (c) to (a);
\draw[edge] (d) to (a);
\foreach \x in {(a), (b), (c), (d)}{ \fill \x circle[radius=2pt];}
\end{tikzpicture}
}

\begin{document}

\title{Fluctuation-dissipation supersymmetry}

\author{Matteo Polettini}

\email{matteo.polettini@uni.lu}
\affiliation{Physics and Materials Science Research Unit, University of Luxembourg, Campus
Limpertsberg, 162a avenue de la Fa\"iencerie, L-1511 Luxembourg (G. D. Luxembourg)} 
 
\date{\today}

\begin{abstract}
Open systems may be perturbed out of equilibrium states either by subjecting them to nonconservative forces or by injecting external currents. For small perturbations, the linear response is quantified by two different matrices. In the framework of network thermodynamics, under very broad assumptions, we show that the two matrices are connected by a supersymmetry, predicting that they have the same spectrum -- up to the degeneracy of the ground state. Our approach brings into the mathematics of supersymmetry a new ingredient, namely oblique projection operators.
\end{abstract} 



\maketitle

\paragraph{Fluctuation-dissipation.} A system can be taken out of equilibrium (a detailed-balance steady state) by
\begin{itemize}
\item[$P$)] subjecting it to nonconservative forces;
\item[$Q$)] injecting external currents.
\end{itemize}
For example, electrical circuits can be plugged into power or current supplies (whence $P$ for power and $Q$ for charge), reaching steady states that violate either of  Kirchhoff's Loop Law (KLL) or Current Law (KCL). In chemical modeling \cite{rao}, one can think of $P$) molecules flowing through a cell's membrane because of osmotic pressure, or $Q$) fixed streams of chemicals into a reactor. In stochastic thermodynamics \cite{broeck}, employing Markov processes to describe dissipation, $P$) occurs when the log-products of the ratios of forward to backward rates along cycles (the {\it cycle forces}) do not all vanish, and the system evolves towards a nonequilibrium steady state; $Q$) corresponds to transients that have not yet reached the steady state, or that undergo a random fluctuation out of it; as they relax back, they produce {\it cocycle currents}.

Far from equilibrium, a system may display dependency on the dynamical activity \cite{baiesi}, negative response \cite{cossetto}, violation of Onsager's symmetry \cite{altaner}, dynamical phase transitions \cite{lazarescu} etc. However, close to equilibrium, the fluctuation-dissipation paradigm states that the response of an (average) current to a external perturbation is not discernible from an internal spontaneous fluctuation (its covariance), and is described by response matrices $H_Q$ and $H_P$, symmetric by microscopic reversibility.

In general, $H_Q$ and $H_P$ will differ, even in rank. However, in this work, employing the formalism of network (or graph) thermodynamics \cite{oster} that encompasses all situations above, we prove that, after making $H_Q$ and $H_P$ suitably adimensional, all eigenvalues larger than $1$ coincide, and that there exists a mapping between the corresponding eigenvectors (response modes), thus proving a spectral fluctuation-dissipation principle. Eigenvalue $1$ may appear, with different degeneracies of the corresponding ground states. In fact, with greater generality we will prove that $H_Q$ and $H_P$ are connected by a supersymmetry (SUSY).

\paragraph{Supersymmetry.} SUSY is the invariance of a theory under a mapping between subspaces of a Hilbert space that entails a degeneracy in the spectrum of its operators. In quantum mechanics, one such Hilbert-space decomposition is between completely symmetric and completely anti-symmetric multi-particle subspaces (bosons and fermions). After second quantization, in relativistic quantum field theory (QFT) the problem of treating bosons and fermions on a similar footing \cite{schwinger,ramon,wess} brought to the development of Grassman variables and graded algebras, that permitted to hypothesise and study a SUSY that mixed the two fields, predicting partner superparticles with the same masses but with opposite spin-statistics. Since no such particle was observed experimentally, it was further speculated that SUSY could be broken, that is, the generator of the symmetry does not annihilate the ground state (in fact, dynamically broken by nonperturbative effects \cite{witten1981}).

SUSY also plays a role as a tool to diagonalize classes of Hamiltonians or find approximation methods in nonrelativistic statistical \cite{nicolai} and quantum \cite{cooper,bagchi} mechanics. Let us review the instructive case of the harmonic oscillator $H = (Q^2 + P^2)/2$, with position $Q$ and momentum $P$ satisfying the canonical commutation relation $[Q,P] = i \hbar I$, with $I$ the identity. Defining the ladder operator $A = (Q + iP)/\sqrt{2}$ one finds $H = A A^\dagger - \frac{\hbar}{2} I = A^\dagger A + \frac{\hbar}{2} I$, $\dagger$ denoting the adjoint. Then, the spectrum of $H$ is easily shown to be $h_0 + \hbar \, \mathbb{N}$, with $h_0$ the ground eigenvalue (to be determined elsewise). This follows from the fact that $AA^\dagger$ and $A^\dagger A$ have the same real nonnegative spectrum up to the multiplicity of eigenvalue zero. \emph{Proof.}  If $\ket{n}$ is an eigenvector of $A^\dagger A$ relative to $n > 0$ then, left-multiplying by $A$, one finds that $A\ket{n}$ is an eigenvector of $AA^\dagger$ relative to the same eigenvalue \emph{$\square$}. (Notice that this argument does not go through for $n =0$ because it might be the case -- as is -- that $A \ket{0} = 0$).

In our proof we will employ similar tools, in particular certain oblique-projection techniques developed by the Author \cite{polettini2015,polettini2020} that involve (possibly non-Hermitean) analogs of $H,Q,P, A$. Before providing the full theory, we discuss a simple but encompassing example.

\paragraph{Example.}

Consider the network in Fig.\,1 with $n=6$ ideal resistors. In Fig.\,2a we depict it as an oriented graph. We label resistances $r_e > 0$ and define matrix $R = \mathrm{diag}\,\{r_e\}$. Ohm's Law states that electric current $j_e$ along a resistor produces a loss of voltage $f_e = r_e j_e$. Let
\begin{align}
\ket{v} := \sqrt{R}^{\,-1} \ket{f} = \sqrt{R} \ket{j}. \label{eq:vdef}
\end{align}
where $\ket{a} = (a_e)_{e}$ denotes a vector, and for a positive-definite matrix $M$, $\sqrt{\,M\,}$ is the unique matrix whose square is $M$. The dissipated power is defined as
\begin{align}
\sigma = \braket{f}{j} = \braket{v}{v} \geq 0 \label{eq:adim}
\end{align}
where $\braket{a}{b} = \sum_{e} a_e b_e$ is the Euclidean scalar product.

Steady currents and potential drops along resistors can be instantaneously produced by plugging into current or voltage sources that provide power ($P$) or charge ($Q$)  (a physical resistor network would also include effective inductors and condensators that smoothen the approach to the steady state). To avoid short-circuits, a maximal number of independent electromotive forces with voltage drops $f_4',f_5',f_5'$ can be placed in parallel to the resistors belonging to a spanning tree $T$ of the network, defined as a maximal set of connections that do not enclose cycles, see Fig.\,2 and Ref.\,\cite{frejtas} for more details on the graph-theoretic architecture of electric circuits. Similarly, a maximal set of independent current generators with input currents $j_1',j_2',j_3'$ can be placed in series to resistors belonging to a cotree $T'_\ast$, defined as a set of connections whose complement is a spanning tree $T'$. While in principle the chosen tree and cotree need not be the complement one of the other (see Fig.\,2c), we will choose $T' = T$, that is, a resistor is either in parallel to a voltage source or in series to a current source, but not both. This non-overlap hypothesis is the only truly restrictive requirement in our formulation. Instead, notice that the converse exercise of plugging a voltage/current generator in series/parallel to a resistor is dull, as it amounts to a base shift of the unknown voltage/current.  

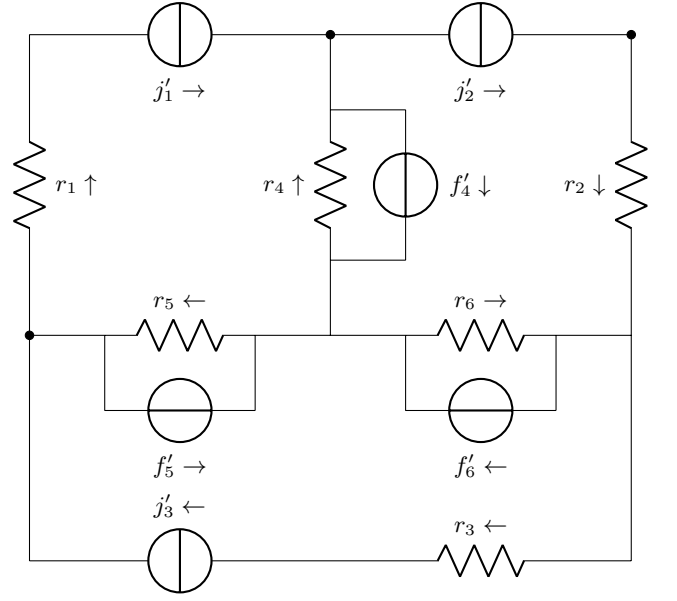
\begin{figure}[t!]
\begin{circuitikz} \draw
(4,6) to (5,6) to[V,l=$f'_4  \downarrow$] (5,4) to (4,4)
(3,3) to (3,2) to[V,l=$f'_5 \rightarrow$] (1,2) to (1,3)
(7,3) to (7,2)  to[V,l=$f'_6 \leftarrow$] (5,2) to (5,3)
(0,7) to[R,  l=$r_1  \uparrow $]  (0,3) 
(4,7) to[I,  l=$j'_1 \rightarrow$]   (0,7)
(4,3) to[R,  l=$r_4  \uparrow $, -*] (4,7)
(0,3) to[R,  l=$r_5 \leftarrow$, *-]    (4,3)
(8,7) to[I,  l=$j'_2 \rightarrow$]  (4,7)
(8,3) to[R,  l=$r_2 \downarrow$, -*] (8,7)
(4,3) to[R,  l=$r_6  \rightarrow$]  (8,3)
(0,0) to (0,3) 
(8,0) to (8,3) 
(0,0) to[I, l=$j'_3 \leftarrow$]
(4,0) to[R,  l=$r_3 \leftarrow$] (8,0)
; 
\end{circuitikz}
\caption{A resistor network powered by a maximal amount of nonoverlapping current and voltage generators. Currents and voltage drops are positive in the direction of the arrows.}
\end{figure}

When all current and voltage generators are connected, the input power is
$\pi = \braket{f'}{j'}$ where $f'_1 = f_1 + f_5 - f_4$, $f'_2 = f_2 + f_4 - f_6$, $f'_3 = f_3 + f_6 - f_5$ are the voltage drops at the extremities of the current generators, obtained using KLL along the $m = 3$ cycles depicted in Fig.\,2d-e-f, and $j'_4 = j_4 + j_1 - j_2$, $j'_5 = j_5 + j_3 - j_1$, $j'_6 = j_6 + j_2 - j_3$ are the currents through the voltage generators, obtained using KCL at the $n-m = 3$  nodes indicated by bullets (i.e. along the cocycles in Fig.\,2g-h-i). Notice that $\pi = \sigma$, that is, all input power is dissipated into the resistors, as per energy conservation (Joule's Law).

The problem to be solved is the determination of the dissipated power when either current or voltage sources are connected. In the first case, obtained e.g. by replacing voltage generators by condensators or open switches, or equivalently by setting  $j'_4 = j'_5 = j'_6 = 0$, we have $j_k = j'_k$ for $k = 1,2,3$. Our objective is to write the dissipation in terms of the rescaled $v_1,v_2,v_3$ only. Using KCL we obtain $\sigma_{Q} = j_1 \left(f_1 + f_4 - f_5 \right) + j_2 \left(f_2 - f_6  - f_4 \right) + j_3 \left(f_3 - f_5 - f_6 \right)$. We now employ Ohm's Law, the KCL again, and rescale. After a few manipulations we find
\begin{align}
\sigma_{Q} = \bra{v} H_{Q} \ket{v} = \bra{v} \left(\ba{cc} H_{Q} & 0 \\ 0 & 0 \ea \right) \ket{v} \label{eq:sigmacur}
\end{align}
with
\begin{align}
H_{Q}  = I_3 + \left(\ba{ccc}
\frac{r_4}{r_1} + \frac{r_5}{r_1} &  - \frac{r_4}{\sqrt{r_1 r_2}}  &  - \frac{r_5}{\sqrt{r_1 r_3}} \\
-\frac{r_4}{\sqrt{r_1 r_2}} & \frac{r_4}{r_2} + \frac{r_6}{r_2}  & -\frac{r_6}{\sqrt{r_2 r_3}} \\
- \frac{r_5}{\sqrt{r_1 r_3}} & -\frac{r_6}{\sqrt{r_2 r_3}} & \frac{r_5}{r_3} + \frac{r_6}{r_3}
 \ea\right).
\end{align}
where $I_m$ is the $m$-dimensional identity matrix (notice that we use the same symbol $H_{Q}$ for the nontrivial block of the full matrix).

Let us now consider the case where the resistor network is only powered by electromotive forces (i.e. the current generators are replaced by inductors or closed switches), i.e. by setting  $f'_1 = f'_2 = f'_6$. Obviously $f_k = f'_k$ for $k = 4,5,6$. Our objective is to write the dissipation in terms of $v_4,v_5,v_6$ only. Performing the same kind of manipulations as above we obtain
\begin{align}
\sigma_{P} & =  \bra{v} H_{P} \ket{v} = \bra{v}  \left(\ba{cc} 0& 0 \\ 0 & H_{P} \ea \right) \ket{v} \label{eq:sigmavol}
\end{align}
where
\begin{align}
H_{P} = I_3 + \left(\ba{ccc} \frac{r_4}{r_1} + \frac{r_4}{r_2} & -\frac{\sqrt{r_4 r_5}}{r_1} & -\frac{\sqrt{r_4 r_6}}{r_2} \\ 
- \frac{\sqrt{r_4 r_5}}{r_1} &  \frac{r_5}{r_1} + \frac{r_5}{r_3} &  -\frac{\sqrt{r_5 r_6}}{r_3} \\
 -\frac{\sqrt{r_4 r_6}}{r_2} & -\frac{\sqrt{r_5 r_6}}{r_3} & \frac{r_6}{r_2} + \frac{r_6}{r_3} \ea  \right).
\end{align}

\begin{figure}[b!]
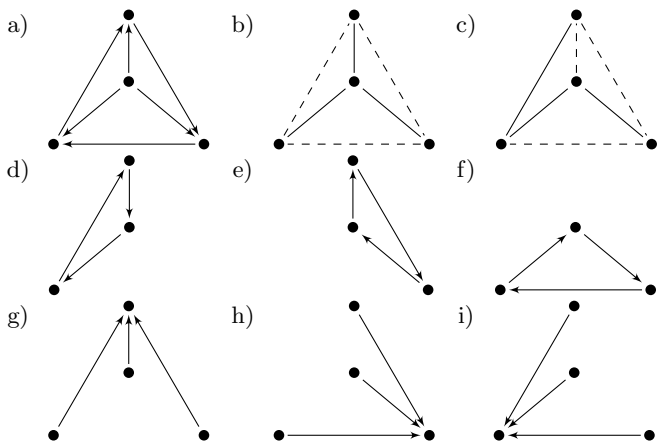


\begin{tabular}{ccc}
a) \networkone[baseline=1.5cm] & 
b) \networktwo[baseline=1.5cm] &
c) \networkthree[baseline=1.5cm] \\
d) \networkfour[baseline=1.5cm] & 
e) \networkfive[baseline=1.5cm] &
f) \networksix[baseline=1.5cm] \\
g) \networkseven[baseline=1.5cm] & 
h) \networkeight[baseline=1.5cm] &
i) \networknine[baseline=1.5cm] 
\end{tabular}

\caption{a) The resistor network as a graph. b) A spanning tree $T$ and, dashed, its complementary cotree $T_\ast$. c) A different choice of tree $T'$ and cotree $T'_\ast$: $T'_\ast$ and $T$ overlap. d-e-f) A basis of cycles generated by $T$. g-h-i) A basis of cocycles generated by $T_\ast$.} 

\end{figure}

The main result of this paper is that the response matrices $H_{P}$ and $H_{Q}$ are supersymmetric in the sense that
\begin{align}
H_{Q} + H_{P} =  AA^\dagger  = \left( \ba{cc} I_m + DD^\dagger & 0 \\ 0 & I_{n-m} + D^\dagger D \ea \right) \label{eq:above}
\end{align}
where here $n=6$, $m=3$, and
\begin{align}
A =  \left(\ba{cc} I_m & -D^\dagger \\ D & I_n \ea\right) \label{eq:above2}
\end{align}
with nonvanishing off-diagonal block
\begin{align}
D =  \left( \ba{ccc} - \sqrt{\frac{r_4}{r_1}} &  + \sqrt{\frac{r_5}{r_1}} & 0 \\
+ \sqrt{\frac{r_4}{r_2}} & 0 & - \sqrt{\frac{r_6}{r_2}} \\
0 & - \sqrt{\frac{r_5}{r_3}} & + \sqrt{\frac{r_6}{r_3}} \ea \right).
\end{align}
This latter matrix is sometimes called the supercharge. It follows that the nonvanishing blocks of $H_P$ and $H_Q$ have the same positive spectrum up to the degeneracy of eigenvalue $1$, as can be checked by direct calculation of the characteristic polynomials.

A few remarks. This particular planar graph is self-dual \cite{servatius} and highly symmetrical. As a consequence, $H_{P}$ can be obtained from $H_{Q}$ by mapping $r_1 \leftrightarrow 1/r_6$, $r_2 \leftrightarrow 1/r_5$ and $r_3 \leftrightarrow 1/r_4$. Also, notice that this particular network admits ground state $\ket{1}$ with all entries $1$, such that $H\ket{1} =\ket{1}$. In generic graphs $D$ will not even be square; then the degeneracy of $1$ will differ. Also, eigenvalue $1$ needs not be present (broken SUSY): it is the case for the self-dual graph $1 \leftrightarrow 2 \leftrightarrow 3 \leftrightarrow 1$ with an additional edge $1\leftrightarrow 2$, which has full-rank supercharge. Finally, the choice of a different spanning tree $T'$ modifies the positive eigenvalues but maintains their overall product (proportional to the weighted spanning (co)tree polynomial \cite{nakanishi}) and the degeneracies of the ground states.

\paragraph*{Network thermodynamics as linear algebra.} To generalize, consider a loopless graph, consisting of vertices connected by edges, arbitrarily oriented from a source to a target. Its topology is encoded in the incidence matrix $\partial$ that maps edges to their boundary vertices. For example the graph in Fig.\,2a) has incidence matrix
\begin{align}
\partial = \left(\ba{cccccc}
-1 &  0 & +1 & +1 &  0 &  0 \\
+1 & -1 &  0 &  0 & +1 &  0 \\
 0 & +1 & -1 &  0 &  0 & +1 \\
 0 &  0 &  0 & -1 & -1 & -1
\ea\right).
\end{align}
The incidence matrix can be seen as an operator acting on the real vector space $\mathcal{E}$ generated by the edges, $\dim \mathcal{E} =n$. Vectors  (currents) in $\mathcal{E}$ are denoted $\ket{j}$, covectors (forces) in the algebraic dual $\mathcal{E}^\ast$ are denoted $\bra{f}$. The action of a covector on a vector $\sigma = \braket{f}{j}$ is called the dissipation. We take $\sigma$ adimensional, so that currents and forces have inverse physical dimensions, and endow $\mathcal{E}$ with a scalar product inducing a canonical isomorphism $\ket{j} \leftrightarrow \bra{j}$ between vectors and forms.

\paragraph*{Kirchhoff's laws and projections.}

We define the cycle space $\mathcal{C} \subset \mathcal{E}$ as the right-null space of $\partial$ and the cocycle space the linear space $\mathcal{C}^\ast_\perp \subseteq \mathcal{E}^\ast$ of covectors that vanish on cycles (the coimage, or row space of $\partial$). Let $m = \dim \mathcal{C}$; therefore $\dim \mathcal{C}^\ast_\perp = n-m$. Cycles and cocycles encode important information about nonequilibrum. Steady states satisfy KCL, $\partial \ket{j} = 0$, therefore steady currents belong to the cycle space $\ket{j} \in \mathcal{C}$. Detailed-balanced systems obey KLL, according to which the net drop of forces along a cycle $\ket{c} \in \mathcal{C}$ is zero, $\braket{f}{c} = 0$. Therefore a detailed-balanced $\bra{f}$ is in the cocycle space $\bra{f} \in \mathcal{C}^\ast_\perp$.

We introduce a projection $P = P^2 : \mathcal{E} \to \mathcal{E}$ onto the space of cycles $P\mathcal{E} = \mathcal{C}$. 
Let $Q = I-P$ be its complement. Since $QP = P - P^2 = 0$, then the coimage of $Q$ is orthogonal to all cycles, and therefore the adjoint $Q^\dagger$ is a projection onto the space of cocycles $Q^\dagger \mathcal{E} = \mathcal{C}_\perp$. Notice that such projections will in general be oblique, that is, not orthogonal, $P^\dagger \neq P$. Expanding the identity we find
\begin{align}
\sigma = \bra{f} (Q + P) \ket{j} = \braket{f_{Q^\dagger}}{j_Q} + \braket{f_{P^\dagger}}{j_P}
\end{align}
where $\bra{f_{P^\dagger}} = \bra{f} P$ (cycle forces) vanish if and only if KLL is satisfied, and $\ket{j_Q} = Q\ket{j}$ (cocycle currents) vanish if and only if KCL is satisfied.

\paragraph*{Linear response, adimensional.}

A linear constitutive relation is a positive operator $R: \mathcal{E} \to \mathcal{E}^\ast$ mapping currents to forces. Notice that in the linear regime the dissipation $\sigma = \bra{j} R \ket{j} \geq 0$ can be seen as a metric. Following Eq.\,(\ref{eq:vdef}), we  introduce adimensional variables. Complementary projections transform like
\begin{align}
P \to \sqrt{R}^{\,-1} P \sqrt{R}, \quad P \to \sqrt{R}^{\,-1} Q \sqrt{R}
\end{align}
(we use the same symbols for the transformed operators). Then KLL and KCL state respectively that $Q \ket{v} = 0$ and $P^\dagger \ket{v} = 0$; we collect these latter fundamental rescaled variables in vector $\ket{v'} = A \ket{v}$, where $A = Q - P^\dagger$ is a skew-symmetric, full-rank operator. Our objective is to write $\sigma = \braket{v}{v}$ in terms of these latter variables. To achieve this, we define
\begin{align}
Q_\perp = Q^\dagger (QQ^\dagger)^+ Q, \qquad
P_\perp = P (P^\dagger P)^+ P^\dagger  
\end{align}
where $M^+$ is the pseudoinverse of $M$.
Importantly, $Q_\perp$ and $P_\perp$ are conjugate orthogonal projections. \emph{Proof.} Using the defining properties of pseudoinverses we find $P_\perp^2 = P_\perp$, $Q_\perp^2 = Q_\perp$, $P_\perp Q_\perp = Q_\perp P_\perp = 0$. Therefore $P_\perp + Q_\perp$ is a full-rank idempotent operator, and only the identity satisfies these requirements \emph{$\square$}. Then we find
\begin{align}
\sigma & = \bra{v} (Q_\perp + P_\perp) \ket{v} = \bra{v'}   H^{-1} \ket{v'}
\end{align}
where
\begin{align}\label{eq:hamiltonians}
H = QQ^\dagger +  P^\dagger P = AA^\dagger
\end{align}
is our desired response operator. Notice that we could write it in terms of $A$, which then plays the role of ladder operator. When KLL is satisfied we have $P \ket{v} = \ket{v}$ yielding $\sigma = \braket{v}{v} = \bra{v} H_P\ket{v}$ recovering Eq.\,(\ref{eq:sigmacur}). Same for KCL and Eq.\,(\ref{eq:sigmavol}). See Ref.\,\cite{polettini2015} for the explicit construction of projection operators from spanning trees.

\paragraph*{SUSY.} It is known \cite{lewko} that complementary projections have the same singular values (eigenvalues of $H_Q$ and $H_P$) but for the multiplicity of $0$ (null states) and $1$ (ground states). To make SUSY manifest, we go in a basis where the response matrices are block diagonal. Then, our main result is the existence of a supercharge $D$ that yields Eqs.\,(\ref{eq:above}) and (\ref{eq:above2}) above.

\emph{Proof.} We decompose our projections in terms of left and right eigenvectors  relative to eigenvalue $1$:
\begin{align}
Q = \sum_{i = 1}^m \ket{q_i^r}\bra{q_i^{\ell}}, \qquad
P = \sum_{j = m+1}^n \ket{p_j^r}\bra{p_j^{\ell}}.
\end{align}
Left eigenvectors are orthogonal to right ones, and can be normalized $\braket{q_i^{\ell}}{q_{i'}^r} = \delta_{ii'}$, for $i,i' \leq m$, and $\braket{p_j^{\ell}}{p_{j'}^r} = \delta_{jj'}$, for $j,j' > m$. Furthermore, since $PQ = QP =0$, we have $\braket{q_i^{\ell}}{p_j^r} = \braket{p_j^{\ell}}{q_i^r} = 0$. In view of Eq.\,(\ref{eq:hamiltonians}), the block-diagonal basis is that of left eigenvectors of $P$ and of right eigenvectors of $Q$. The matrix entries are found by evaluating onto the dual basis, which, without loss of generality can be chosen to be orthonormal, $\braket{q_i^{\ell}}{q_{i'}^{\ell}} = \delta_{ii'}$ and $\braket{p_j^r}{p_{j'}^r} = \delta_{jj'}$, since both span degenerate eigenspaces. We obtain
\begin{align}
(H_Q)_{ii'} & = \bra{q_i^{\ell}} Q^\dagger Q \ket{q_{i'}^{\ell}} = \braket{q_i^r}{q_{i'}^r} \\
(H_P)_{jj'} & = \bra{p_j^r} PP^\dagger \ket{p_{j'}^r} = \braket{p_j^{\ell}}{p_{j'}^{\ell}}.
\end{align}
Also, in this basis $A$ takes the form of Eq.\,(\ref{eq:above2}), and with supercharge $D_{ij} = \braket{q_i^{\ell}}{p_j^{\ell}} = - \braket{q_i^r}{p_j^r}$, where this latter identity, crucial and nontrivial, is obtained by expanding  $0 = \braket{p^r_i}{q^{\ell}_j} = \bra{p^r_i} (Q+P) \ket{q^{\ell}_j}$. The right-hand side of Eq.\,(\ref{eq:above}) can then be checked by direct calculation \emph{$\square$}. 

\paragraph{Conclusions.}

In this work we proved that the adimensionalized linear-response  matrices obtained by perturbing equilibrium states either by $P$) nonconservative forces, or by $Q$) injection of currents, are connected by a SUSY that predicts the same spectrum, but for the degeneracy of the ground state. SUSY also predicts relations between the eigenvectors (modes of response) that deserve further investigation. To illustrate the result we employed a highly symmetrical resistor network, plugged into a maximal number of nonoverlapping current/voltage sources. However, the theory applies to any discretized thermodynamic system in the linear regime, in particular to chemical networks and stochastic thermodynamics, and to generic electrical circuits.

The cycles/cocycles analysis clearly relates to concepts in differential geometry.  The connection between the cohomology of graphs and electrical circuits dates back to Weyl \cite{weyl}, who also introduced a method to solve circuits in terms of orthogonal projections \cite{bamberg}. Witten \cite{witten} connected cohomology to SUSY-QFT via the analogy with Laplacian operator $DD^\dagger + D^\dagger D$ (where now $D$ is the exterior differential and $\dagger$ the Hodge dual). For graphs, operator $H = QQ^\dagger + P^\dagger P$ is clearly reminiscent of a Laplacian, but it is to be noticed that the theory we outlined does not necessarily require any graph-theoretical (``spatial'') origin of the projection operators.

Let us put forward another argument suggesting a deeper connection. In SUSY-QFT, the supercharge maps fermions to bosons and vice versa. In our formalism, matrix $D$ maps left-$Q$ and right-$P$ eigenvectors, that is, (the generating edges of) cycles to cocycles, and viceversa. This analogy can be backed up by another argument. A path $\ket{\gamma}$ on a network can be characterized by the net number of times $n_Q(i) = \braket{q_i^\ell}{\gamma}$ and $n_P(j) = \braket{p_j^\ell}{\gamma}$ that cocycle $i$ or cycle $j$ are performed. By the path's continuity, cycles can occur an arbitrary number of times (in either direction), $n_P(j) \in \mathbb{Z}$,  but cocycles can only be crossed at most once, $n_Q(i) \in \{-1,0,1\}$. These are reminiscent of the occupation numbers for bosons/fermions.

To conclude, while we do not foresee immediate applications of our results -- the program of unification of the forces under SUSY is stalled due to the lack of a perspective of an experimental validation, and nonequilibrium thermodynamics is mostly interested in systems far from the linear regime -- there are several areas where it is necessary or convenient to consider projections onto physically meaningful states that are not orthogonal -- such as molecular orbitals \cite{mulliken,manning}, coherent states, non-commuting observers \cite{ivanovic,peres} -- and where our methods might apply.

\paragraph*{Acknowledgments.} 

The research was supported by the National Research Fund Luxembourg (project CORE ThermoComp R-AGR-3425-10) and by the European Research Council, project NanoThermo (ERC-2015-CoG Agreement No. 681456).

\end{document}